\documentclass[review]{elsarticle}
\usepackage{graphicx}
 \usepackage[usenames]{color}
\usepackage{lineno,hyperref}
\usepackage{lineno}
\usepackage{geometry}
\journal{JINST}

\newcommand{\fer}[0]{FNAL-NICADD}
\newcommand{\elj}[0]{ELJEN EJ-208}
\newcommand{\prot}[0]{PROTVINO}

\begin{document}

\begin{frontmatter}
\title{{\bf Characterisation of plastic scintillator paddles and \\ lightweight MWPCs for the MID subsystem of ALICE 3}}

\author[if_unam]{Ruben~Alfaro}
\author[icn_unam]{Mauricio~Alvarado~Hern\'andez}
\author[wigner]{Gyula~Benc\'edi}
\author[itn]{Juan~Carlos~Cabanillas~Noris}
\author[icn_unam]{Marco~Antonio~D\'iaz~Maldonado}
\author[uas]{Carlos~Duarte~Galvan}
\author[fcfm_buap]{Arturo~Fern\'{a}ndez~T\'{e}llez}
\author[wigner]{Gergely~G\'abor~Barnaf\"{o}ldi}
\author[wigner]{\'Ad\'am Gera}
\author[if_unam]{Varlen~Grabsky}
\author[wigner]{Gerg\H{o}~Hamar}
\author[cinvestav]{Gerardo~Herrera~Corral}
\author[uas]{Ildefonso~Le\'on~Monz\'on}
\author[fcfm_buap]{Josu\'e~Mart\'inez~Garc\'ia} 
\author[fcfm_buap]{Mario~Iv\'an~Mart\'inez~Hernandez}
\author[icn_unam]{Jes\'us~Eduardo~Mu\~noz~M\'endez}
\author[wigner]{Rich\'ard Nagy}
\author[itn]{Rafael~\'Angel~Narcio~Laveaga}
\author[icn_unam]{Antonio~Ortiz}
\author[fcfm_buap]{Mario~Rodr\'iguez-Cahuantzi}
\author[prague]{Solangel~Rojas~Torres} 
\author[prague]{Timea~Szollosova}
\author[icn_unam]{Miguel~Enrique~Pati\~no~Salazar}
\author[icn_unam]{Jared~Pazar\'an~Garc\'ia}
\author[fcfm_buap]{Hector~David~Regules~Medel}
\author[fcfm_buap]{Guillermo~Tejeda~Mu\~noz}
\author[icn_unam]{Paola~Vargas~Torres}
\author[wigner]{Dezs{\H o}~Varga}
\author[wigner]{R\'obert~V\'ertesi}
\author[fcfm_buap]{Yael~Antonio~Vasquez~Beltran}
\author[icn_unam]{Carlos~Rafael~V\'azquez~Villamar}
\author[fcfm_buap]{Irandheny~Yoval~Pozos}

\address[if_unam]{Instituto de F\'isica, Universidad Nacional Aut\'onoma de M\'exico, Mexico}
\address[icn_unam]{Instituto de Ciencias Nucleares, Universidad Nacional Aut\'onoma de M\'exico, Mexico}
\address[wigner]{HUN-REN Wigner Research Centre for Physics, Hungary}
\address[itn]{Instituto Tecnol\'ogico de Culiac\'an, Tecnol\'ogico Nacional de M\'exico}
\address[uas]{Universidad Aut\'onoma de Sinaloa, Mexico}
\address[fcfm_buap]{Facultad de Ciencias F\'isico Matem\'aticas, Benem\'erita Universidad Aut\'onoma de Puebla, Mexico}
\address[cinvestav]{Centro de Investigaci\'on y Estudios Avanzados del IPN, Mexico}
\address[prague]{Czech Technical University in Prague, Prague, Czech Republic}


\cortext[mycorrespondingauthor]{antonio.ortiz@nucleares.unam.mx}


\begin{abstract}

The ALICE collaboration is proposing a completely new detector, ALICE~3, for operation during the LHC Runs 5 and 6. One of the ALICE~3 subsystems is the Muon IDentifier detector (MID), which has to be optimised to be efficient for the reconstruction of $J/\psi$ at rest (muons down to $p_{\rm T}\approx1.5$\,GeV/$c$) for $|\eta|<1.3$. Given the  modest particle flux expected in the MID of a few Hz/cm$^2$, technologies like plastic scintillator bars ($\approx1$\,m length) equipped with wavelength-shifting fibers and silicon photomultiplier readout, and  lightweight Multi-Wire Proportional Chambers (MWPCs) are under investigation. To this end, different plastic scintillator paddles and MWPCs were studied at the CERN T10 test beam facility. This paper reports on the performance of the scintillator prototypes tested at different beam momenta (from 0.5\,GeV/$c$ up to 6\,GeV/$c$) and positions (horizontal, vertical, and angular scans). The MWPCs were tested at different momenta (from 0.5\,GeV/$c$ to 10\,GeV/$c$)  and beam intensities, their efficiency and position resolutions were verified beyond the particle rates expected with the MID in ALICE 3. 

\end{abstract}

\begin{keyword}
\texttt{Upgrade,} \texttt{Scintillators,} \texttt{SiPM,} \texttt{MWPC,} \texttt{ALICE experiment} 
\end{keyword}

\end{frontmatter}

\section{Introduction
\label{sec:Introduction}}

ALICE (A Large Ion Collider Experiment) is one of the four main experiments at the CERN Large Hadron Collider (LHC) and it is mainly devoted to heavy-ion physics~\cite{Aamodt:2008zz}. Its main goal is to achieve the detailed characterisation of the strongly interacting quark--gluon plasma (QGP) that is theoretically predicted in lattice QCD, and that has been observed in relativistic heavy-ion collisions by experiments at RHIC and at the LHC~\cite{Busza:2018rrf,Bala:2016hlf}. 

In spite of the ambitious upgrade and physics programs to be completed in the next eight years, crucial questions will remain open after the ALICE data taking in the LHC Runs 3 and 4. These include the deeper understanding of the rich phenomenology of QCD matter by connecting parton-energy loss, collective flow, hadronisation and electromagnetic radiation in a unified description. The successful completion of this task requires the construction of a new detector with increased rate capabilities, improved vertexing and tracking over wide momentum and rapidity ranges. This project, named ALICE~3, is being optimized to have excellent capabilities for heavy-flavour and electromagnetic radiation studies~\cite{ALICE:2022wwr}. The ALICE 3 detector is expected to take data during the LHC Run~5 (2035-2038) and Run~6 (2040-2041). The core of ALICE~3 will be a full silicon tracking device covering eight units of pseudorapidity, a Time-Of-Flight (TOF) system consisting of two cylindrical layers with a time resolution of 20\,ps, a Ring Imaging Cherenkov (RICH) detector that will allow ${\rm e}/\pi$ separation up to 2\,GeV/$c$, and finally an electromagnetic calorimeter covering the barrel acceptance and one forward direction. All these detectors will be located inside a superconducting magnet, providing a magnetic field of $B = 2$\,T. A Forward Conversion Tracker (FCT) is under consideration  to measure photons down to transverse momenta of $p_{\rm T}\approx2$\,MeV/$c$. For muon identification, the Muon IDentifier (MID) detector will consist of a cylindrical shape steel absorber of thickness $\approx70$\,cm, surrounding the magnet, and of MID chambers around the absorber allowing the muon tagging by matching hits in the muon detectors with tracks from the tracker. The absorber thickness is being optimized for the efficient reconstruction of $J/\psi$ emitted at rest at $\eta=0$, which requires the muon identification with a transverse momentum as low as $p_{\rm T}=1.5$\,GeV/$c$.

Regarding the MID subsystem, muon chambers with a granularity of $\Delta\varphi\Delta\eta = 0.02 \times 0.02$ are proposed~\cite{ALICE:2022wwr}. The moderate charged-particle rate in the MID subsystem of a few Hz/cm$^{2}$ for pp and Pb--Pb collisions at an interaction rate of 24\,MHz and 93\,kHz, respectively,  together with the required granularity of 50-60\,mm pad size open the possibility of using well known technologies like Multi-Wire Proportional Chambers (MWPCs) and Resistive Plate Chambers (RPCs). However, the use of plastic scintillator bars equipped with wave-length shifting fibers (WLS) and silicon photomultiplier (SiPM) readout, is considered as the baseline option in the ALICE~3 LoI due to their simplicity~\cite{ALICE:2022wwr}. This option considers two layers of crossed scintillator paddles (5\,cm wide) with a 10\,cm gap between two layers. Both MWPC and scintillators represent a cost effective solution to cover a detection area of about $360$\,m$^{2}$.   

In this paper, the performance of different options of plastic scintillator bars (FNAL-NICADD~\cite{Pla-Dalmau:2000puk}, ELJEN model EJ-208~\cite{refEljen} and U-70 IHEP, PROTVINO) and photosensors are reported.  For MWPCs, two sets of chambers with different sizes and wire geometries ($512\times512\,{\rm mm}^2$, wire spacing: \,8mm, pad: 4\,mm) and ($768\times768\,{\rm mm}^{2}$, wire spacing: 12\,mm, pad: 12\,mm) are tested with high and moderate position resolution, measuring their performance at beam conditions, as well as quantifying the performance in high-particle rate environment.

\section{Experimental setup
\label{sec:ExperimentalSetup}}

\begin{figure}[ht]
\centering
\includegraphics[width=0.95\linewidth]{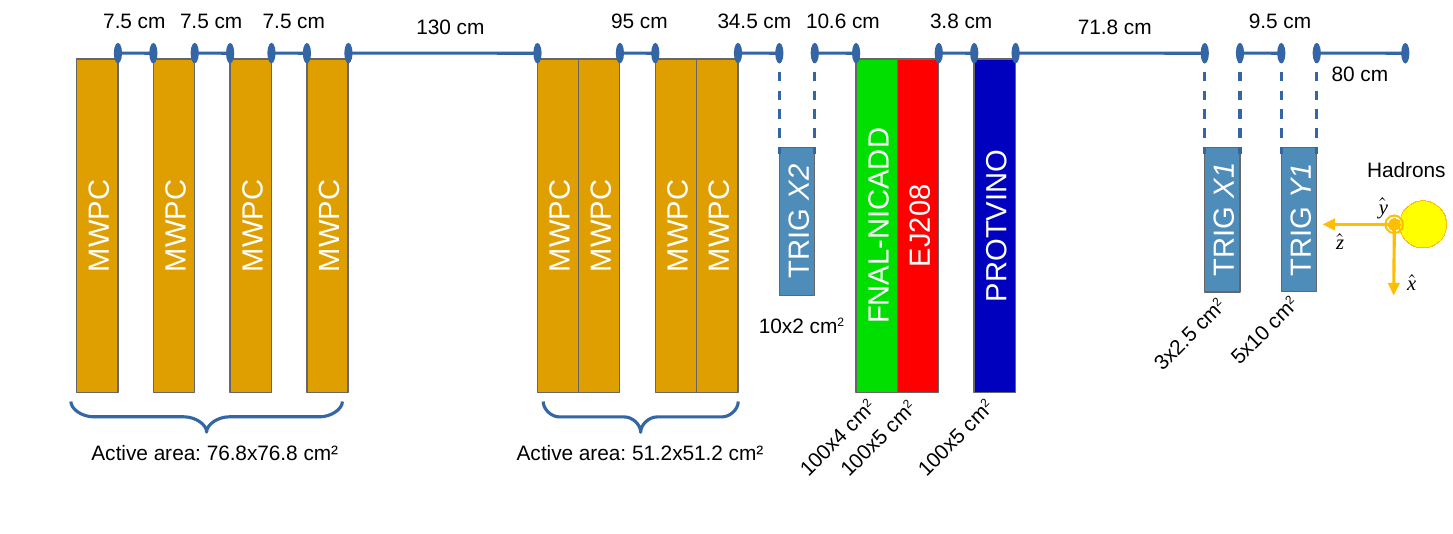}
\caption{Sketch of the experimental setup. The trigger scintillators are labeled as TRIG.~$X1$, TRIG.~$Y1$ and TRIG.~$X2$. The beam travels from right to left along the $z$ coordinate that is perpendicular to the array of detectors tested at T10. The detector areas are given in terms of the length along the horizontal axis ($x$) times the length along the vertical axis ($y$).}
\label{fig:expsetup}
\end{figure}

The measurements were performed at the T10 beamline at the East Area of the CERN PS, which underwent a recent overhaul~\cite{Bernhard:2020}, allowing a convenient beam testing activity.  Three scintillator prototypes and two sets of MWPCs were tested. Figure~\ref{fig:expsetup} shows a sketch of the experimental setup and the coordinate system. The beam particles travel along the $z$ axis, the horizontal axis mentioned hereinafter corresponds to the $x$ axis, whereas the vertical axis corresponds to the $y$ axis. The trigger signal was provided by the coincidence among three scintillator paddles (hereinafter called trigger scintillators) denoted as TRIG.~$X1$, TRIG.~$Y1$ and TRIG.~$X2$. The dimensions of the paddles are indicated in Figure~\ref{fig:expsetup}. The thickness of TRIG.~$X1$ was 2.5\,cm, whereas the other two were 1\,cm thick. The trigger scintillators are coupled to Hamamatsu R3478 photomultipliers with short lightguide, resulting in high efficiency and a high signal-to-noise ratio with a time resolution of about 1~ns. 

The MWPCs operated in a self-trigger mode, which, despite its limited time resolution, facilitated the detection of all particles from the beam due to their expansive coverage area. The detected particles include those that missed the trigger scintillators due to their small detection area ($\approx 3\times2$\,cm$^{2}$). Thus, the real beam profile could be measured.

\subsection{Plastic scintillator prototypes}

The low-cost extruded plastic scintillator bar from FNAL-NICADD~\cite{Pla-Dalmau:2000puk} was included among the prototypes. Other prototypes based on the relatively low-cost scintillator manufactured by ELJEN (model EJ-208) and plastic scintillator manufactured at U-70 IHEP, \prot\, (which was used for the cosmic-ray detector of ALICE~\cite{Fernandez:957722}) were also tested. All of them had approximately one meter in length. The \fer\, and \elj\, scintillator prototypes were constructed and tested with cosmic muons at UNAM, and the \prot\, scintillator prototype was constructed and tested at BUAP.  Table~\ref{tab:1} summarises the main features of the plastic scintillator bars and the model of the SiPMs used in the prototypes.  A brief description of the plastic scintillator prototypes can be found below. 

\begin{table}[h!]
\centering
\caption{Characteristics of the plastic scintillator bars. The width (w), length (l) and thickness (t) are specified, as well as the number of fibers and SiPM model.}
\begin{tabular}{c | c | c | c | c}
 \hline\hline
   {\bf Company } & {\bf Bar dimensions}  & {\bf No. of fibers} & {\bf Fiber diameter} & {\bf SiPM model} \\ [0.25ex]
    & {\bf  (${\rm w} \times {\rm l} \times {\rm t}$) } &  &  &  \\ [0.5ex]
 \hline
 \fer & $4.0\times100.0\times1.0\,{\rm cm}^{3}$ &  1 & 1.5\,mm & Hamamatsu S13360-6050CS \\
 \elj & $5.0\times100.0\times1.0\,{\rm cm}^{3}$ &  0 & -- & Hamamatsu S13360-6050CS \\
 \prot & $5.0\times100.0\times1.0\,{\rm cm}^{3}$ & 1  & 1.5\,mm &  SensL serie C $6\times6\,{\rm mm}^2$ \\
 \hline\hline
\end{tabular}
\label{tab:1}
\end{table}

Given the relatively large area to be covered, an attractive alternative is the low-cost scintillator produced at the FNAL-NICADD facility~\cite{Pla-Dalmau:2000puk}, which has been used in several experiments (see e.g. Ref.~\cite{McFarland:2006pz,MICHAEL2008190}). This plastic scintillator has a light emission peak at 420\,nm~\cite{Grachov:2004jg}. Since the attenuation length of the plastic scintillator is short (a few tens of cm), the light produced by the particle interaction has to be collected, reemitted, and transported to the photodetectors by wave-length shifting (WLS) fibers. The prototype used the Kuraray WLS fiber model Y-11(200) that poses a long-attenuation length ($\approx400$\,cm) with an emission (absorption) peak at 476\,nm (375\,nm)~\cite{kuraray:ref}.  The fiber is mounted inside the $2$\,mm groove on the large surface of the scintillator coupled only with air. A Hamamatsu SiPM of type S13360-6050CS was coupled at one side of the fiber mentioned above. The SiPM has a photosensitive area of $6.0\times6.0$\,mm$^{2}$, pixel pitch of 50\,$\mu$m, its detection spectral response is found in the wavelength interval 270-900\,nm, with a photon detection efficiency of 40\% at peak sensitivity wavelength (450\,nm). The SiPM was connected to the module SP5600 manufactured by CAEN, which serves to provide the bias voltage (57\,V) and an analog output connector. 

Another alternative that was tested is the relatively low-cost plastic scintillator produced by ELJEN (EJ-208). This scintillator has a long attenuation length of about 4\,m and therefore no WLS fibers were considered in this prototype. This plastic scintillator has a light emission peak at 435\,nm with rise (decay) time of 1\,ns (3.3\,ns). The coupling with the SiPM (S13360-6050CS) was done with an optical grease (ELJEN) having high transparency and a refractive index close to the refractive index of the plastic scintillator ($n=1.58$). The CAEN module SP5600  was also used in this prototype that serves to provide the bias voltage (57\,V) and an analog output connector. 

The \prot\, plastic scintillator bar was equipped with two SiPMs at each side of the bar and one Kuraray WLS fiber model Y-11(200). Two SiPMs were needed otherwise the efficiency at a large distance with respect to SiPM was observed to drop down to 80\%. Data for beam momentum of 6 GeV/c were used. The SiPMs used in this prototype were produced by SensL (C-series) with an active area of $6.0\times6.0$\,mm$^{2}$ operated at $-30$\,V. The fiber was directly connected to the SiPMs without any optical grease. Its photon detection efficiency is about 40\% at peak sensitivity wavelength ($\approx420$\,nm). A custom-made front-end electronics card was used. The front-end electronic circuit was designed and manufactured to comply with the specifications outlined in the SIPM datasheet provided by the manufacturer~\cite{refOnsemi} and the requirements of the detectors. The circuit uses only resistors and capacitors to facilitate both biassing and readout of the SiPM.

Data acquisition was done using standard VME and NIM modules. The three triggers scintillation, TRIG.~$X1$, TRIG.~$Y1$, and TRIG.~$X2$ in Fig.~\ref{fig:expsetup}, were connected to a NIM leading edge discriminator (CAEN N840). Two of the signals were processed with a coincidence module (LeCroy 622A) to define the two-fold coincidence. The two-fold coincidence was defined by TRIG.~$X1$ $\wedge$ TRIG.~$Y1$, while the three-fold one was defined by the coincidence among  TRIG.~$X1$ $\wedge$ TRIG.~$Y1$ $\wedge$ TRIG.~$X2$.

The two-fold coincidence signal served as the primary trigger to initiate data acquisition, while the third trigger signal was digitised, and the three-fold coincidence was done during the offline analysis. 
The signals from the SiPMs were fully digitised using the CAEN V1742 digitiser with 32 channels, enabling the simultaneous recording of signals from all prototypes. The digitiser was configured with a sampling rate of 1\,GS/$s$. A TTI~PL330DP power supply provided the voltage (30\,V) to SensL SiPMs. The output signals from these SiPMs were amplified tenfold before being connected to the digitiser using a variable fast amplifier (CAEN V974).  The data of the 32 channels were transferred to a computer with the VME bridge (CAEN V1718) for further processing and analysis using standard CAEN software.

\subsection{MWPC prototypes}

The MID subsystem requires high efficiency ($\approx100\%$) and lightweight detectors. The so-called ``CCC''  (closed-cathode chamber~\cite{CCC_2013}) MWPC used for similar purposes in the VHMPID upgrade project of ALICE~\cite{VHMPID} has been investigated in the past. The purpose of the present beam test was to verify the compatibility of similar MWPC structures with the MID requirements. For this reason, a more standard but still lightweight and low-cost MWPC using contemporary materials and design concept~\cite{VCI_2020} was built. Similar MWPC detectors are broadly applied for cosmic muon imaging, where high reliability and robust structures are required~\cite{AHEP_2016}. 

Eight MWPC detectors spaced over more than 3\,m length, defining very precise individual beam particles, were tested. Two detector sizes were used, labeled as ``50\,cm'' and ``80\,cm'' prototypes. The detectors have a gas gap of 18--22\,mm, an anode wire spacing of 8\,mm for ``50\,cm'' and 12 mm for ``80\,cm'' prototypes, with two-dimensional projective readout achieved by field wires parallel to the anode wires, as well as perpendicular ``pick-up'' wires~\cite{AHEP_2016}. These pick-up wires function effectively the same as the cathode strips, or as the ``pads'' of the CCC-design. Whenever a particle crosses the chamber, the anode wire signal is capacitively coupled to the neighbouring electrodes: typically two (sometimes three) field wires, and two to four pick-up wires give signal above an appropriately set threshold.
This results in a pair of hit clusters for both directions.

Having 8 MWPCs provided sufficiently precise and redundant track information. A ``track'' is defined if at least 7 out of the 8 available detectors form a straight line within the position resolution of the detectors. Individual detector efficiencies are typically above 95\%, which ensures high efficiency for track identification.

The detector system was controlled and data were read out using a DAQ system designed for cosmic muons~\cite{VCI_2020}, however, it proved to be sufficiently fast to operate up to the highest relevant beam intensity. The dead time of the DAQ system was about 0.1\,ms, leading to 10\,kHz maximum event rate for recording.

\begin{figure}[ht]
\centering
\includegraphics[width=0.75\linewidth]{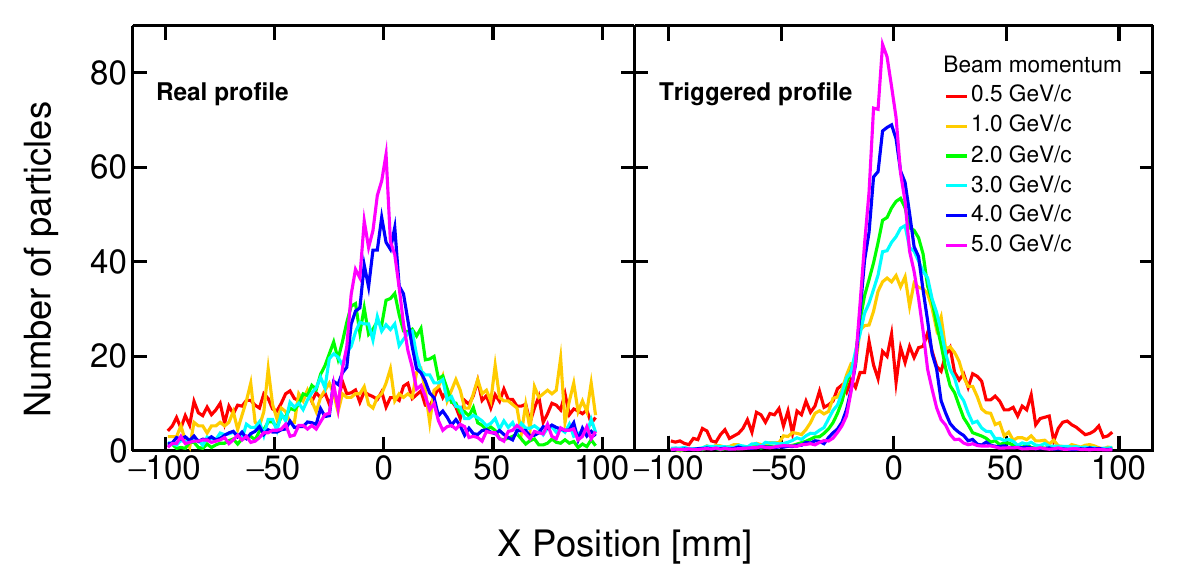}
\caption{Horizontal beam profiles (number of particles along the $x$ axis) measured with MWPCs at various beam momenta. The beam profile is shown considering the self-triggered MWPC mode  (left panel) and triggered mode using the trigger scintillators (right panel).}
\label{fig:beamprofiles}
\end{figure}

The beam profile at any plane perpendicular to the beam direction was determined extrapolating the individual tracks to the beam source. The beam profile was found to be strongly energy and position dependent (along the \(z\) axis), and slightly intensity dependent. Figure~\ref{fig:beamprofiles} shows an example of the $x$ position of the first MWPC. The left panel shows the beam profile (number of particles along the $x$ axis) with triggering only on the MWPC detectors, which is the ``real'' beam profile reaching the detectors. The right-hand side panel shows that the two-fold coincidence signal used for efficiency measurements gives a much cleaner beam definition. Apparently, the beam widens at lower energies as expected~\cite{Bernhard:2020}, but it is conveniently usable for the measurements over the full energy range. The triggered beam profile is relevant for efficiency measurements, however the complete beam profile must be considered when the beam intensity is determined at any specific detector location.

\section{Results
\label{sec:Results}}

\subsection{Plastic scintillator bars
\label{sec:Results-Scintillators}}

\begin{figure}[ht]
\centering
\includegraphics[width=0.95\linewidth]{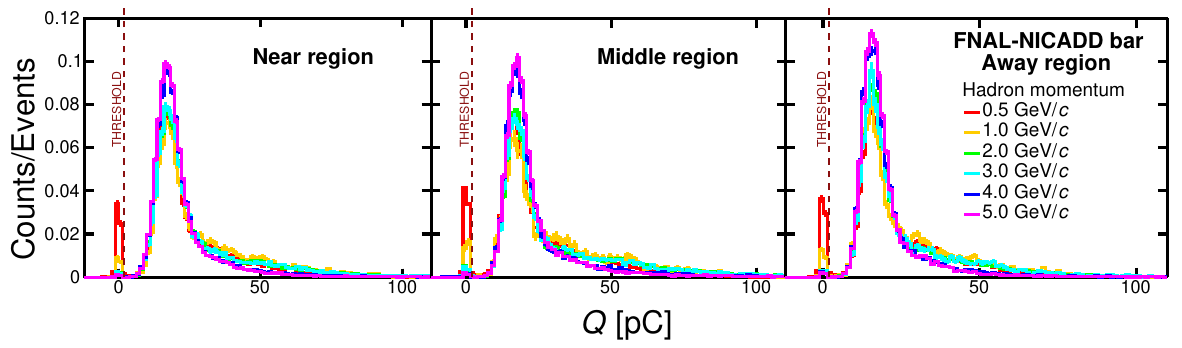}
\caption{Charge distributions measured for the \fer\, prototype. Results are displayed for different beam momenta, as well as for three different distances between the hit and the SiPM of $\approx5$\,cm (left), $\approx50$\,cm (middle), and $\approx95$\,cm (right).}
\label{fig:1}
\end{figure}

To illustrate the quality of data, the charge ($Q$) distributions obtained with the \fer\, plastic scintillator prototype are shown in Fig.~\ref{fig:1} for different beam momenta as well as for different distances between the hit and the SiPM. The charge was obtained by integrating the pulse within a time interval of 300\,ns. The time was measured relative to the arrival time of the trigger signal. The main peak of the signal exhibits little or no variation with respect to the distance due to the WLS fiber. A model based on a convolution of a Landau and a Gaussian distribution is fitted to the charge distributions. The mean of the Gaussian and the most-probable value (MPV) of the the Landau function are constrained to be the same. The Gaussian distribution serves to describe detector effects that smear the charge distribution.  The measured  charge (MPV) is around 18\,pC that corresponds to $\approx40$ photoelectrons. This value was obtained following a standard technique for single photon calibration~\cite{ CHMILL201770}.
Figure~\ref{fig:1} shows a clear and consistent signal-to-pedestal separation as a function of hit position and beam momentum. A similar good pedestal/signal separation is seen for \elj\, and \prot\, prototypes. Although \elj\, scintillator prototype gives a light output of up to 160\, photoelectrons, the main disadvantage of this prototype is that the plastic scintillator cost is 4 times that of \fer.  
To study the effect of the integration time window, its size was varied from 200\,ns to 350\,ns keeping the start of the time window fixed (with respect to the time arrival of the trigger signal). The MPV of the charge variation is below 10\% and is stable with the beam momentum.

\begin{figure}[ht]
\centering
\includegraphics[width=0.55\linewidth]{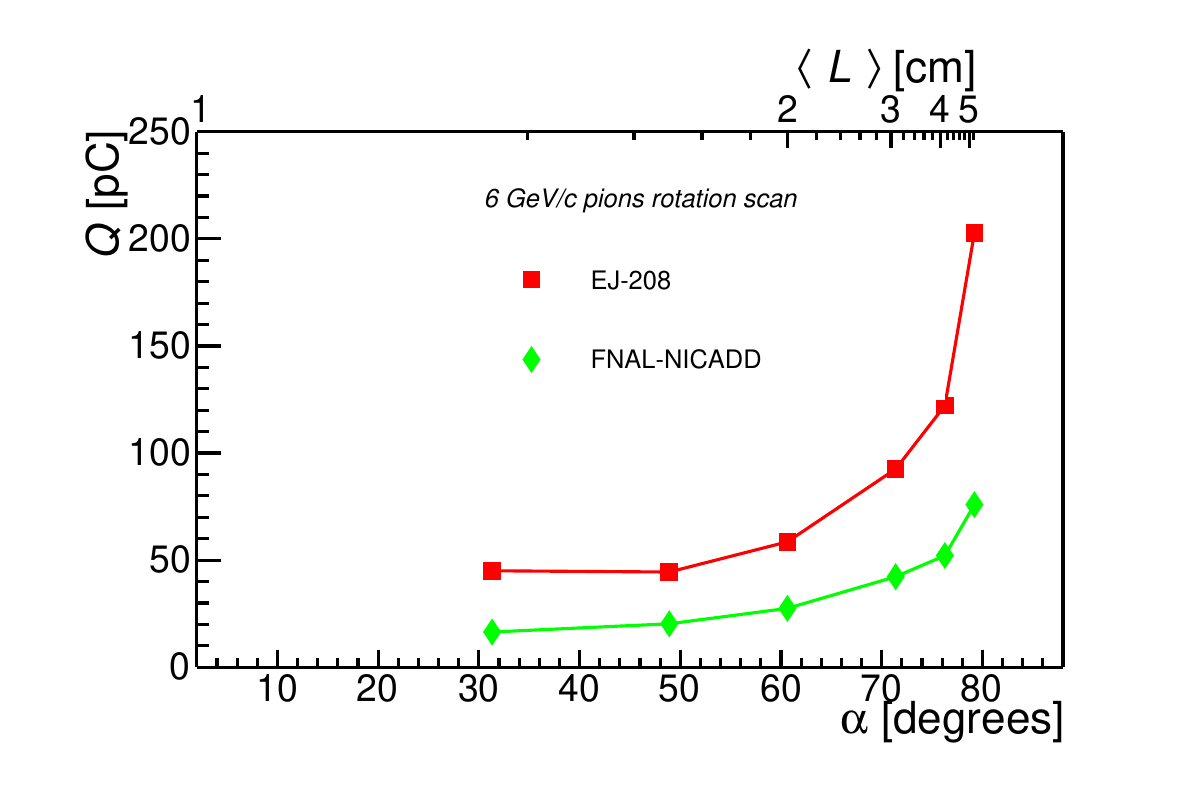}
\caption{Charge (MPV) as a function of the incidence angle and path length ($\langle L \rangle$). Results for \fer\, and \elj\, are displayed. The statistical errors are smaller than the markers and, therefore, are not visible.}
\label{fig:2.1}
\end{figure}

The light collection is expected to depend on the mean path length ($\langle L \rangle$) traveled by the particle in the scintillator bar. The data presented in this paper consider hadrons crossing the scintillator bars at normal incidence (incidence angle of $\sim 0^{\circ}$), therefore they traverse the 1\,cm thickness of the bar. In experiments at colliders like LHC, only particles coming from the interaction point at pseudorapidity equal to zero ($\eta=0$) would satisfy this condition. However, the acceptance of the MID subsystem is expected to be around $|\eta|<1.24$, therefore particles at $\eta=1.24$ (incidence angle of $\sim 58^{\circ}$) might reach the detector. Particles reaching MID at larger incidence angles ($60-90^{\circ}$) originated from beam-induced background could also be observed. It is, therefore, useful to study the detector response as a function of the incidence angle. This study was conducted by rotating the prototypes covering the interval between $30^{\circ}$ and $80^{\circ}$. Figure~\ref{fig:2.1} shows the charge (MPV) as a function of the incidence angle and path length. As expected, the charge value shows a modest variation ($\sim 1.7$) for incidence angles within  $30^{\circ}-60^{\circ}$, which is roughly the interval covered by particles from the interaction point. For smaller angles, the charge exhibits a steep increase with decreasing the incidence angle. For angles of about  $\sim 80^{\circ}$, the charge is four times that achieved at normal incidence.

\begin{figure}[ht]
\centering
\includegraphics[width=0.45\linewidth]{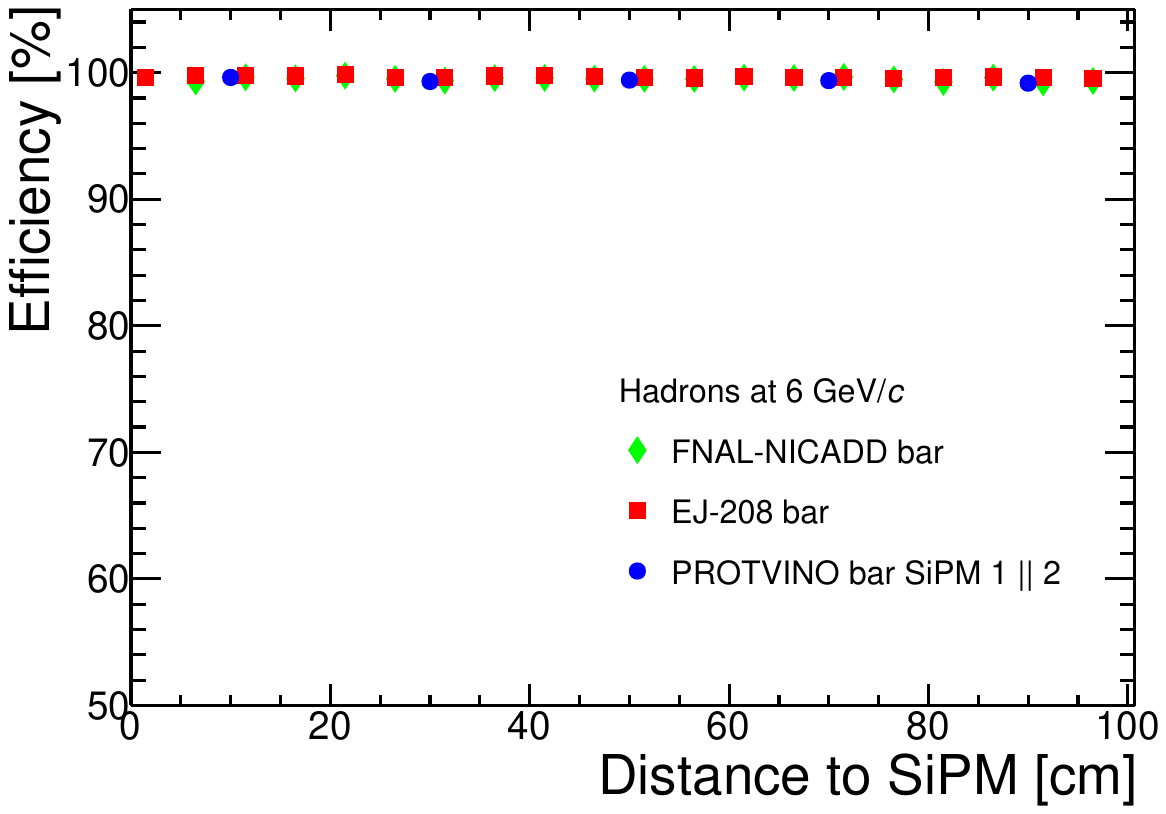}
\includegraphics[width=0.45\linewidth]{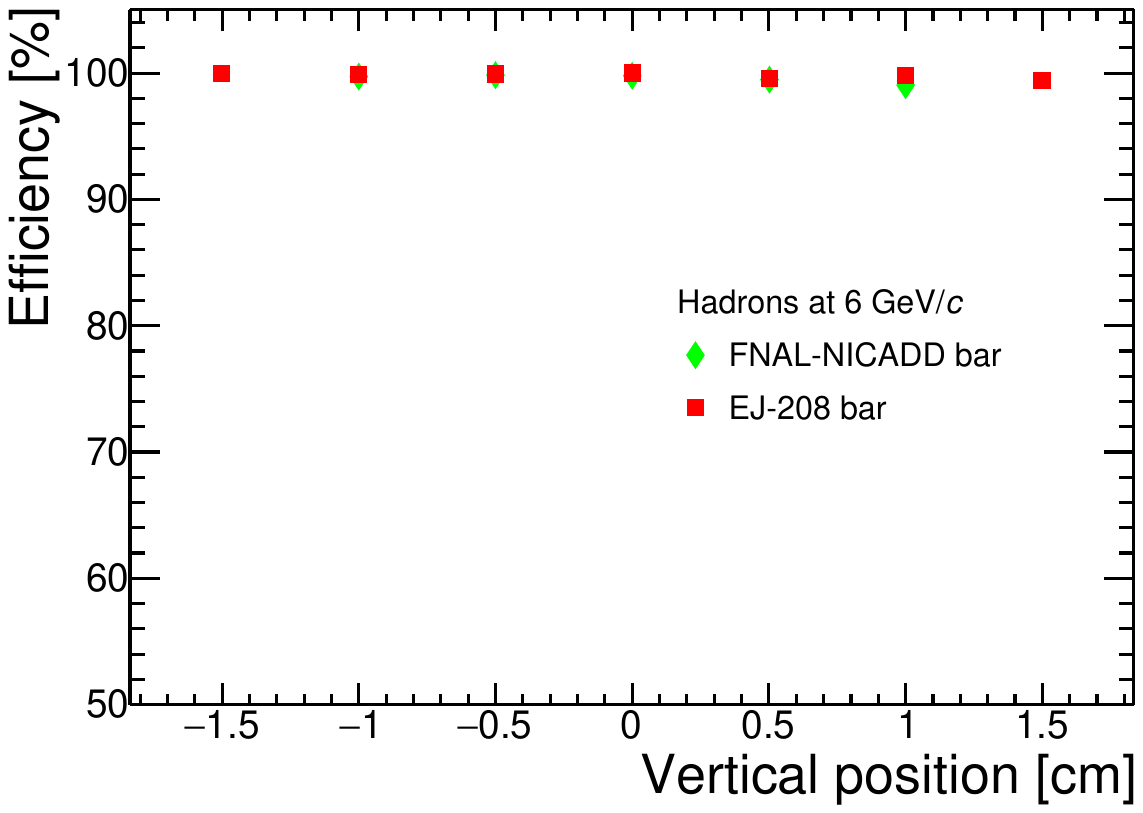}
\caption{Left: efficiency as a function of the distance between the SiPM and the hit position (horizontal scan). Right: efficiency along the width of the bar (vertical scan). The statistical errors are smaller than the markers and, therefore, are not visible.}
\label{fig:4}
\end{figure}

The efficiency is determined by considering events that are triggered by the triple coincidence TRIG. $X1$ $\wedge$ TRIG. $Y1$ $\wedge$ TRIG. $X2$. It is obtained as the ratio of the number of events registered by the prototype (measured charge above the threshold) divided by the total number of events. In the case of \prot\, prototype, the detector signal comes from any of the two SiPMs placed at each side of the bar.  The efficiencies as a function of the hit position along the bar length are shown in Fig.~\ref{fig:4}. Results for hadron momentum of 6\,GeV/$c$ are presented. The efficiencies are above 98\% along the length of the bar, and the detector efficiency is found to be quite uniform. For completeness, a similar measurement was performed along the bar width (vertical scan) at mid position (hit position at 50\,cm from the SiPM). Given the size of the active area of the trigger, $2\times2$\,cm$^{2}$, the efficiency measurement was restricted to $\pm1.00$\,cm ($\pm1.5$\,cm) around the center of the scintillating bar for \fer\, (\elj) prototype. The right panel of Fig.~\ref{fig:4} shows that within those intervals, the efficiency is above 98\% for both prototypes.

The time resolution of scintillator prototypes is mostly determined by the length of the scintillator bar ($\approx1$\,m) and is about 1.7\,ns and 1.5\,ns for \fer\, and \elj\, prototypes, respectively. This time resolution is already enough to operate the detector in the expected LHC conditions during Runs 5 and 6. In the case of ALICE 3 the expected interaction rate for pp collisions is of around 24\,MHz. Assuming an integration time of the ALICE~3  tracker of about 100\,ns~\cite{ALICE:2022wwr}, a pileup of around 2-4 pp collisions is expected. The measured time resolution in our prototypes would be more than enough for the MID subsystem.

\subsection{MWPC
\label{sec:Results-MWPC}}

Similar ``CCC-type'' detectors were thoroughly characterised in test beams earlier~\cite{CCC_2013} and field measurements for muon imaging~\cite{gi-1-229-2012}. Detailed studies on MWPC with cosmic muons have been reported~\cite{VCI_2020}, therefore this paper primarily addresses the intensity dependence of the MWPC performance.


The beam test conditions allowed the measurements to span over a broad range of beam intensities by adjusting the collimator settings and by changing the position along the diverging beamline. Particularly, beam rates two orders of magnitude above the expected rate in the ALICE~3 MID could be safely reached during the tests.

Intensity is defined as particle rate divided by the beam spot area. The beam area could be determined assuming a Gaussian beam profile,  $A=\pi \sigma_x \sigma_y$ where $\sigma$ is the width of the beam spot. The beam profile is shown in Fig.~\ref{fig:beamprofiles} at the respective position of the detector under study. The particle rate was derived directly from the count rate measured using default beam-instruments of T10 and our DAQ. The estimation of the intensity has a considerable systematic error, around  30\% as averaging on a mostly-Gaussian distribution, to which the main contribution is from the non-Gaussian shape of the beam. 

\begin{figure}[ht]
\centering
\includegraphics[width=0.5\linewidth]{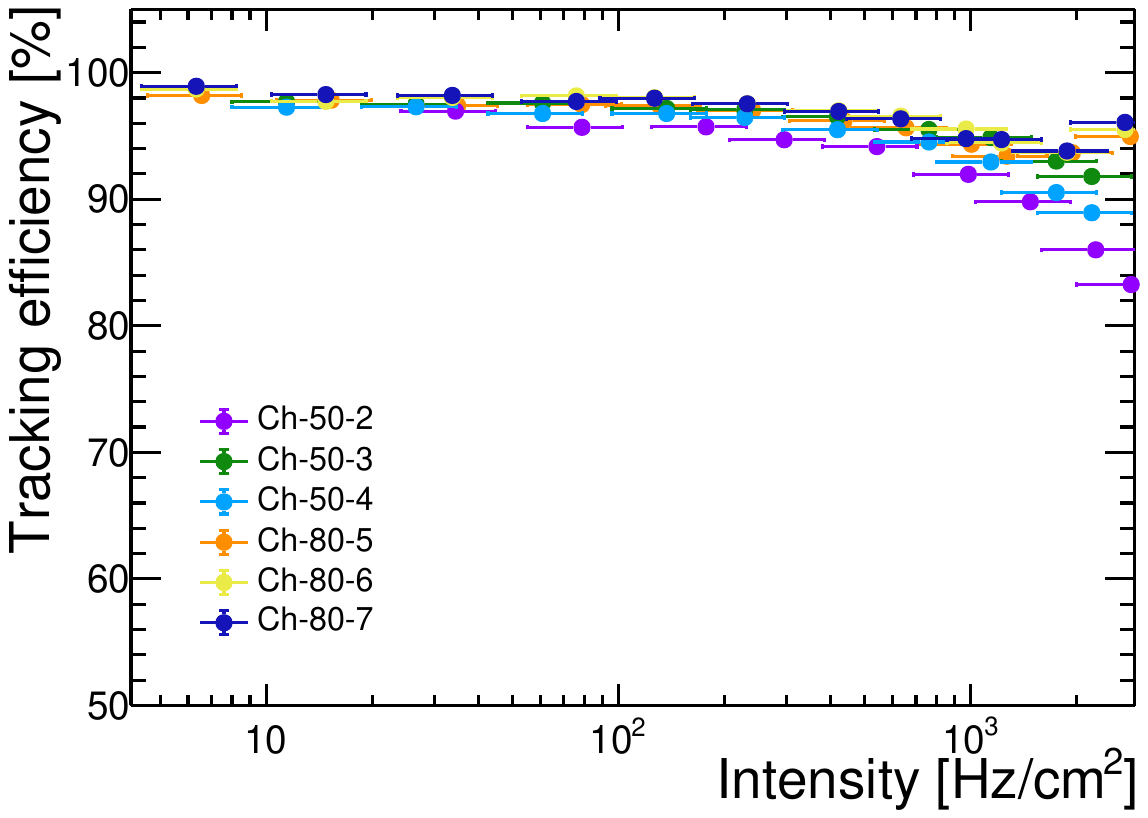}
\caption{MWPC detection efficiency measured as a function of beam intensity for beam momentum of 5\,GeV/$c$.}
\label{fig:eff_intensity_mwpc}
\end{figure}

Figure~\ref{fig:eff_intensity_mwpc} shows the measured efficiency as a function of the estimated beam intensity. The efficiency is defined as the probability to find a valid hit in a given chamber, under the condition that using the other 7 chambers, a straight line track can be formed using at least 6 chambers.

The measured efficiency is around 98\% at beam intensity of 10\,Hz/cm$^2$, and decreases only slightly until a steep drop around 1\,kHz/cm$^2$. Efficiency drop is expected from various sources \cite{SAULI}: a physical limit is set by the shielding effect of ions around the anode wires, expected to be relevant at much higher rates. In our case, since the top beam intensities were saturating the DAQ with a total beam rate well above 10\,kHz/cm$^2$, it is possible that the efficiency drop is connected to the readout electronics rather than the detector amplification structure. The key conclusion is that the efficiency is above 95\% and constant for intensities up to 100 Hz/cm$^2$, well above the intensities expected at the ALICE~3 MID.


The position information from the MWPC chambers is acquired from wires running parallel and perpendicular to the anode wires, called ``pads'' and ``field'' wires, respectively \cite{AHEP_2016}. In order to simplify the readout system, only a discriminated signal was registered from each wire, which results in a position resolution related to the segmentation (8\,mm and 12\,mm for the 50\,cm and 80\,cm versions, respectively). The segmentation is considerably better than that of the scintillator bars ($5/\sqrt{12}$\,cm), which seems sufficient for the purpose of the MID detector.

The measurement of position resolution in a specific chamber uses the information provided by the eight layers considered in the setup. When a track is identified (with at least seven valid hits along a straight line), the chamber under study is removed from the tracking, the track is further fitted considering the remaining (six or seven) hit points, and then the difference between the measured point in the chamber under study, and the track intersection is determined.

\begin{figure}[ht]
\centering
\includegraphics[width=0.5\linewidth]{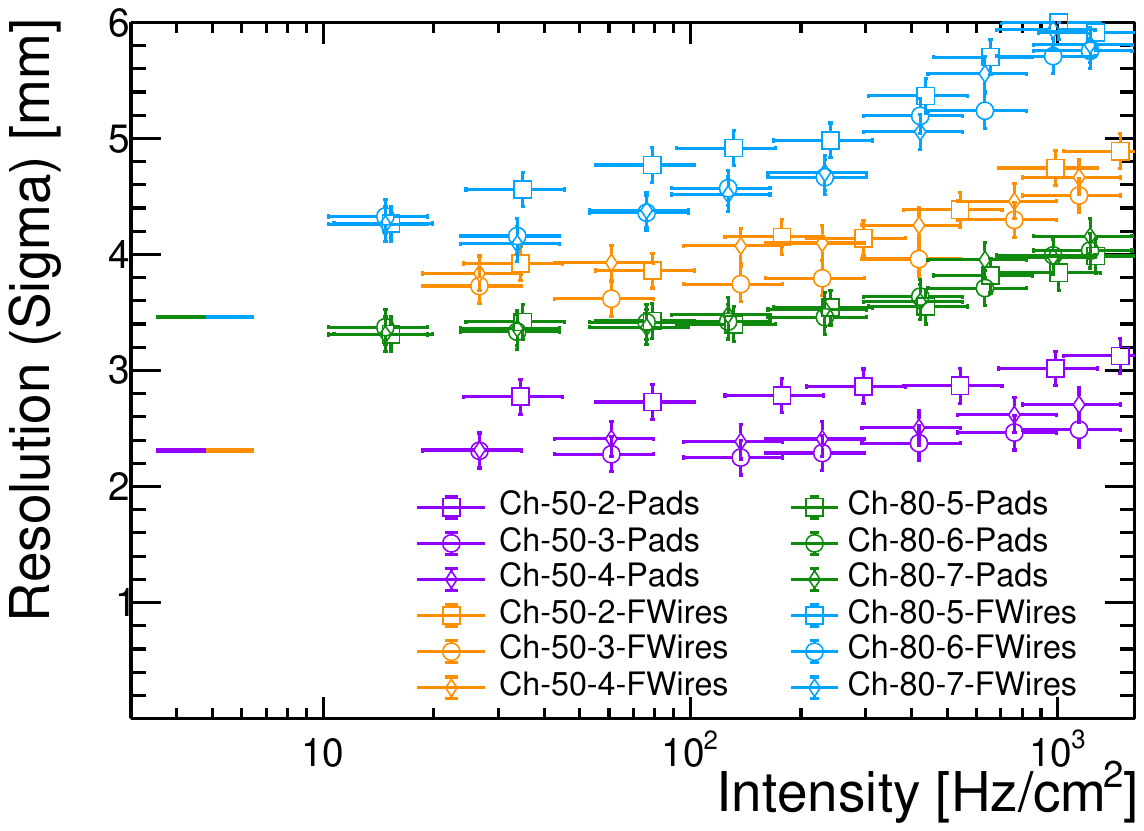}
\caption{MWPC position resolution as a function of beam intensity. Measurements were performed for beam momentum of 5\,GeV/$c$. The short horizontal lines correspond to the resulution expected for 12mm (8mm) segmentation of the 80cm (50cm) chambers.}
\label{fig:res_intensity_mwpc}
\end{figure}

The measured position resolution is expected to degrade at high intensities. Figure~\ref{fig:res_intensity_mwpc} shows the measured resolution as a function of the beam intensity. There is a clear broadening of the resolution toward high intensities. The resolution is about 10-30\% worse for intensities around 1000\,Hz/cm$^2$, where the efficiency drop was observed. It is interesting to compare the measured resolution with the expectation $\sigma_0$ from the known segmentation. These values, $12/\sqrt{12}$ mm and $8/\sqrt{12}$ mm are indicated in the figure. Due to the complicated signal formation in the chambers, measured resolution of ``pads'' and ``field wires'' is different. In addition, the resolution was measured using information from other chambers. These systematic effects do not change with intensity.


\begin{figure}[ht]
\centering
\includegraphics[width=0.5\linewidth]{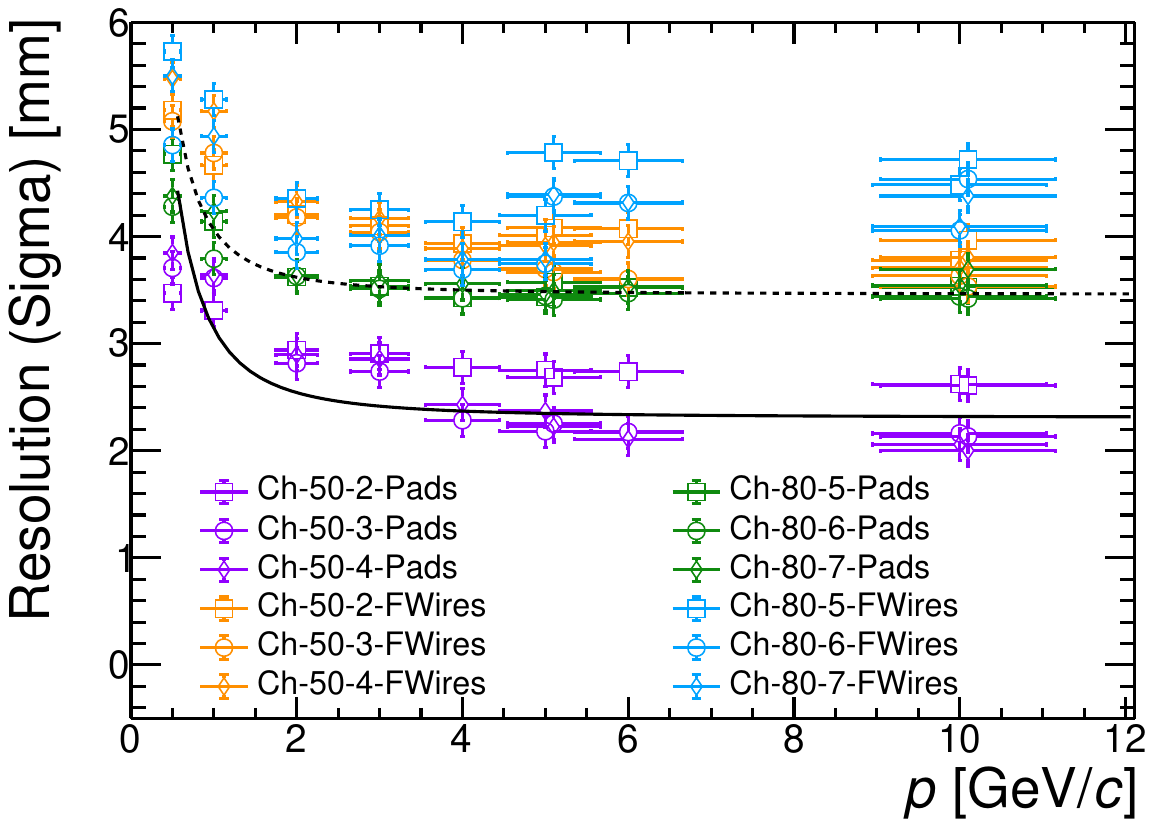}
\caption{MWPC position resolution as a function of beam momentum. The lines correspond to the quadratic sum of a constant term and an $S/p$ scattering term (see text for explanation). Here $S$ is set to 2.125\,mm$({\rm GeV}/c)^{-1}$.}
\label{fig:res_energy_mwpc}
\end{figure}

Regarding the momentum dependence, one can expect the degradation of the apparent resolution due to the scattering of the particles along the beam line. 
Figure~\ref{fig:res_energy_mwpc} shows the measured position resolution as a function of beam momentum. Particularly at the lowest momenta, 0.5 and 1\,GeV/$c$, the broadening amounts to about 40-60\%. Assuming that scattering contributes to the broadening of a momentum-independent intrinsic momentum resolution, the data can be parametrised as $\sigma(p)=\sqrt{\sigma_0^2+(S/p)^2}$, where $\sigma_0$ is the scattering-free limiting value, which is around 2.31 and 3.46 for the 8 mm and 12 mm segmentation, respectively. This function, as indicated in Figure~\ref{fig:res_energy_mwpc}, shows the correct tendency, but does not allow the precise determination of the parameter $S$.  Considering a rough estimate for $S$ (obtained from the fits),  $S \approx $ 2.125\,mm$({\rm GeV}/c)^{-1}$, one can extract the material budget of the experimental setup. Considering the detector geometry (three groups of chambers define the $L=2.5$\,m trajectory) the $S$ parameter can be estimated as $S \approx (L/2) (14\,{\rm MeV}) \sqrt{x} (1+0.038\ln(x))$,  where $x$ is expressed in units of radiation length. This formula yields $x \approx 2\%$ which matches well the actual material budget of about 5\,mm glass epoxy.


The systematic measurements presented above indicate that up to beam intensities of 100\,Hz/cm$^2$, similarly to what was observed for the efficiency, there is a very small degradation of position resolution and that the angular spread of the beam particles does not dominate the measured resolution above $p=2$\,GeV/$c$.



\section{Summary and outlook \label{sec:Conclusions}}
The ALICE~3 experiment will entail a large-area muon detector, with high and uniform efficiency, and excellent time resolution, and running with moderate occupancy and particle fluence. Scintillator, MWPC, and RPC based options are all taken into account, and two of these technologies were tested at the CERN T10 test beam facility.

Regarding plastic scintillator prototypes, the option that offers good performance on light-yield output (around 40 photoelectrons), good time resolution ($<2$\,ns), and that represents a moderate-cost solution is the extruded \fer\, scintillator bar equipped with WLS fiber and only one SiPM at one side of the bar. Other possibilities were tested, in particular \elj\, scintillator bar also exhibits a good performance but the cost per bar is 4 times that of the \fer\, option. For the \prot\, scintillator bar to be competitive, it had to be equipped with WLS fiber and two SiPMs. However, having two sensors per bar in the MID detector would increase the cost of the readout electronics. The other issue is that the plastic is more than 40 years old and its technical specifications are missing. 

Concerning the tested MWPCs, the detectors were functioning reliably during the test beam period, and the results offer new detailed information with control on beam intensity, beam energy, and geometry, which matches the MID structure. The tested MWPC type is competitive, having high efficiency and excellent position resolution even beyond the required particle fluence.  Its moderate intrinsic time resolution shall be checked against the expected rate and occupancy in ALICE~3 MID simulations.


\appendix


\section*{Acknowledgments}

Support for this work has been received from CONAHCyT under the Grants CF No. 2042, CB A1-S-22917 and A1-S-13525; PAPIIT-UNAM under the project IG100524; VIEP-BUAP 2023/213; from Ministry of Education, Youth, and Sports of the Czech Republic, grant number LM2023040; as well as from Hungarian National Research, Development and Innovation Office under the Grants: OTKA-FK-135349, ELKH-KT-SA-88/2021, TKP2021-NTKA-10, and 2021-4.1.2-NEMZ-KI-2022-00018/11223-23. Detector construction used the VLAB facilities at HUN-REN Wigner RCP. This project has received funding from the European Unions Horizon Europe research and innovation programme under grant agreement No 101057511. Authors acknowledge the ALICE Collaboration and T10 CERN team for facilities, software and organisation of the test beam.


\bibliography{ref}

\end{document}